# CASHEW DATASET GENERATION USING AUGMENTATION AND RaLSGAN AND A TRANSFER LEARNING BASED TINYML APPROACH TOWARDS DISEASE DETECTION

Varsha Jayaprakash[1], Akilesh K[1], Ajay kumar[1], Balamurugan M.S [2,*] Manoj Kumar Rajagopal[2,*]

[1,2] School of Electronics Engineering, VIT Chennai, 600127, India;
varsha.jayaprakash2019@vitstudent.ac.in

*Email: balamurugan.ms@vit.ac.in

**ABSTRACT:**

Cashew is one of the most extensively consumed nuts in the world, and it is also known as a cash crop. A tree may generate a substantial yield in a few months and has a lifetime of around 70 to 80 years. Yet, in addition to the benefits, there are certain constraints to its cultivation. With the exception of parasites and algae, anthracnose is the most common disease affecting trees. When it comes to cashew, the dense structure of the tree makes it difficult to diagnose the disease with ease compared to short crops. Hence, we present a dataset that exclusively consists of healthy and diseased cashew leaves and fruits. The dataset is authenticated by adding RGB color transformation to highlight diseased regions, photometric and geometric augmentations, and RaLSGAN to enlarge the initial collection of images and boost performance in real-time situations when working with a constrained dataset. Further, transfer learning is used to test the classification efficiency of the dataset using algorithms such as MobileNet and Inception. TensorFlow lite is utilized to develop these algorithms for disease diagnosis utilizing drones in real-time. Several post-training optimization strategies are utilized, and their memory size is compared. They have proven their effectiveness by delivering high accuracy (up to 99%) and a decrease in memory and latency, making them ideal for use in applications with limited resources.

Keywords: Cashew, anthracnose, image processing, augmentation, RaLSGAN, transfer learning, tinyML.

## 1. INTRODUCTION

Agriculture production is a significant contributor to the Indian economy. India is one of the world's major cashew producers. The country produces over 600,000 tonnes of cashews per year, accounting for approximately 30% of worldwide production [1]. The largest cashew-producing states in India are Kerala, Andhra Pradesh, and Tamil Nadu. Many small farmers in these states rely on cashew growing for a living. According to a recent study, overall crop loss in India amounts for about 15-25% of total output [2,3]. They can be caused by a variety of issues such as pests and fungal diseases, poor management, acidity of soil owing to overtreatment, and poor treatment. Among these, the most widely infected disease that affects the yield of cashew trees is anthracnose. It's one of the deadliest phytosanitary problems that affects most of the tropical fruits worldwide. Anthracnose is caused by

Colletotrichum gloeosporioides that is a fungal pathogen that attacks the leaves, nuts, and peduncles of cashew plants.[4]. The disease spreads by causing the plant's color to change to orangish brown or dark brown, depending on the age of the fungus. The interior seed tissues and seedlings might get infected with the disease. [5]. Manual examination by farmers or professionals has been the traditional method of disease identification, but this method may be time-consuming and expensive, making it impracticable for millions of small and medium-sized farms worldwide [6]. As the population of the earth rises, finding ways for detecting and mitigating plant diseases has the twin objective of improving agricultural productivity and lowering pesticide use.

Development of automated systems for identifying plant diseases based on visible indicators on leaves has been accelerated by the development of computer vision models [7]. These innovations aim to make farmer participation as easy, feasible and the detection system as reliable as possible. This may be accomplished by employing deep learning algorithms that have been trained on big datasets containing images of healthy and diseased cashew trees. One of the primary advantages of employing computer vision to detect cashew disease is that it can be done rapidly and correctly with minimum human intervention. This can save time and money while also increasing the speed and accuracy of disease diagnosis. As many resources are not available, a dataset that contains images of healthy and diseased cashew leaves and fruits is created. This dataset contains images that are a combination of early infections and highly infected ones, with images taken at different heights, brightness, and time of the day.

Further this dataset can be tested with tinyML algorithms and employed in drones using cameras that can fly over huge regions of cashew crops and photograph them [8,9]. Drones and unmanned aerial vehicles (UAVs) may fly over huge fields and snap photographs of the plants from numerous perspectives, providing a full view of the crop. The tinyML algorithms may then be used to evaluate the photos and identify disease symptoms such as leaf discoloration or wilting. It may be used to more accurately anticipate diseases that the farmer may not be able to detect. It also eliminates the need to transport data to a cloud server, which may necessitate an IoT connection, making the overall system more efficient by increasing speed, reducing computing strain, and operating with low-power hardware devices [10-12]. Furthermore, it may be integrated with a control system that can take appropriate steps to eliminate infected regions or utilize fertilizers based on the output of the classifier system.

In this paper, a dataset exclusively for cashew leaves and fruits is presented. It contains 20,127 images of healthy and anthracnose infected images. Few tinyML techniques such as MobileNet and Inception are applied using images of cashew leaves and fruits to develop a binary classifier that can distinguish between healthy and anthracnose-infected leaves and test the prepared dataset. The dataset is further expanded by utilizing RaLSGAN (Relativistic Average Least Squares Generative Adversarial Network) to improve the authenticity and depth of study. The paper is organized as follows: Section II describes the existing resources and methodologies to generate and test the data, Section III and Section IV explains the dataset and the methodology adopted to create it, Section V describes the evaluation methodology using CNN and transfer learning, Section VI discussed the results obtained followed by future works and conclusion.

## 2. LITERATURE SURVEY

A dataset to detect diseases in black gram leaf is presented by Talasila et.al. [13]. The dataset consists of 1000 images of leaves. A total of 5 classes are present out of which 4 are diseased and 1 class is healthy. It consists images of diseases like anthracnose, Leaf Crinkle, Powdery Mildew and Yellow Mosaic. The data is acquired in the coastal region of Andhra Pradesh using high quality DSLR cameras and mobile phones. The data is then pre-processed and augmented using rotation, noise injection, illumination etc. to expand the dataset for further classification. Later, this data is used in a CNN model like Alexnet, GoogleNet, Mobilenet and Resnet. Results proved that this data is beneficial in classifying the disease with an average accuracy of 98%.

Rauf et. Al. [14] present a dataset exclusively for citrus leaves and fruit disease classification. The various diseases are sub categorized as black spot, scab, greening, melanose and canker. The dataset consists of 759 images of both healthy and infected leaves and fruits, with 609 images of eaves and 150 images of fruits. The self-annotated images are then processed and converted to 256 x 256 pixels, filtered using gaussian function and top hat filter to improve contrast, and are segmented using weighted segmentation and saliency maps. Finally, the color and geometric features are extracted from the image such that it can be used for future classification purposes.

Zhao et al. [15] have proposed a method that utilizes Double GAN to generate images of unhealthy plant leaves, thus balancing plant datasets. This is achieved through a two-step process. In the first step, healthy leaf images were used to create a pre-trained model using Wasserstein generative adversarial network. Unhealthy leaf images were then used with the pre-trained model to generate 64x64 pixel images. In the second step, a super-resolution generative adversarial network was employed to transform the 64x64 pixel images into 256x256 pixel images to expand the dataset. The generated images were compared to those produced by DCGAN, and it was observed that the images generated by Double GAN were much clearer. Furthermore, the accuracy in identifying plant species and diseases was higher in the expanded dataset than in the original dataset.

Xinda Liu, et al [16] have introduced a new, extensive dataset of plant diseases, consisting of 271 plant categories and over 220,000 images. The objective is to solve the difficulties in visual plant disease detection for precise diagnosis of plant diseases, in particular, due to the random distribution of lesions, the wide range of symptoms, and the complicated backdrops. The suggested strategy involves reweighting both visual areas and loss to emphasize diseased sections in order to improve disease detection. This method also utilizes the LSTM network to represent features, resulting in better performance compared to other approaches.

Gianni Fenu et al. [17] have developed and shared the DiaMos plant field dataset, which includes 3505 images of pear fruits and leaves with four different levels of severity for each of the four-leaf diseases. The dataset also covers four stages of fruit growth, and the images were captured using both smartphones and DSLR cameras. The images are presented in two different resolutions, 2976 x 3968 and 3456 x 5184, and were taken in real-world scenarios under different lighting conditions and angles. This dataset provides valuable guidelines for constructing and selecting datasets for future research in this field.

Barbole et al. [18] present the RGB dataset and RGB-D dataset are two sub-datasets that make up the GrapesNet dataset. 5,000 handheld camera photos of grapevines are included in the RGB dataset. Each photograph has a size of 640x480 pixels and was taken at various phases of grapevine development. 2,500 photos of grapevines were taken with a handheld camera and a depth sensor and are included in the RGB-D collection. The depth sensor records depth data in addition to color data, which improves comprehension of the grapevines' three-dimensional (3D) structure. The RGB-D graphics have a 640x480 pixel resolution. Using a deep learning model trained on the GrapesNet dataset, it demonstrated an average accuracy of 95.5% on a binary classification test of healthy and sick grapevines.

Geon Heo et al. [19] discuss few data collecting methodologies like data sources, data labelling, data cleansing, and data augmentation. Also, the authors talk about the difficulties and advantages of gathering data and offer suggestions for future studies in the area. Large volumes of data may be processed and managed with the aid of Big Data technologies like Hadoop and Spark. Data labelling and cleaning can benefit from the use of AI techniques like computer vision (CV) and natural language processing (NLP). The authors also discovered that methods for enhancing data, such data synthesis and picture manipulation, may significantly enhance the effectiveness of machine learning models.

Shorten et al [20] discuss about numerous picture data augmentation methods, broken down into fundamental image alterations like kernel filters and random erasing. By combining neural style, mixing images, and geometric transformation with deep learning techniques like adversarial training, neural style transfer, and GAN data augmentation, as well as mixing images, color space transformation, and geometric transformation, a category of meta learning known as neural augmentation is produced. This survey goes into great depth about the use of auto-augment and smart-augment approaches.

Perez et al. [21] ran tests on two datasets, CIFAR-10 and CIFAR-100, each of which contains 60,000 32x32 color pictures of 10 and 100 classes, respectively, to assess the efficacy of data augmentation. A baseline model, a model with data augmentation, and a model with dropout regularization were the three deep learning models they trained. The outcomes demonstrated that data augmentation greatly increased the models' accuracy on both datasets, lowering the error rate by as much as 50%. Moreover, it was shown that combining data augmentation with dropout regularization increased the models' accuracy even more.

In order to properly diagnose and treat diseases that may affect crop yields and quality, Thapa et. al. [22] presents a work by underlining the significance of plant pathology, particularly in the context of agriculture. Apple scab and cedar apple rust are two of the most prevalent foliar diseases in apple trees, according to the authors, and they can cause serious harm if left untreated. Over 40,000 photos of both healthy and unhealthy apple leaves make up the dataset they have created. Expert plant pathologists analyzed the photographs after gathering them from different American orchards. Images of apple scab, cedar apple rust, healthy leaves, and leaves with many additional illnesses are all included in the dataset. They examined the performance of several machine learning models, including convolutional neural networks (CNNs), on the dataset in order to assess the dataset and the classification findings. With an accuracy of over 95% on the test set, they discovered that CNNs performed the best.

While deep learning and transfer learning techniques have shown promise for illness classification, Barbedo et al. [23] describe research that shows that their effectiveness is dependent on a variety of factors, including dataset size and diversity. Over 54,000 pictures of 14 different plant diseases made up the dataset they utilized for their studies, which is described in the article. The photographs are collected from a variety of places, including web databases, field research, and lab tests. To train and evaluate several deep learning and transfer learning models, the scientists separated the dataset into three subgroups of variable sizes and disease variety. On the three separate datasets, they compared the performance of several models, including a pre-trained CNN, a fine-tuned CNN, and a transfer learning model. They discovered that model performance increased as dataset size and variety increased, with the transfer learning model obtaining the greatest accuracy of over 99% on the biggest and most varied dataset.

Parraga-Alava et. al. [24] have created a dataset with more than 9,000 pictures of both healthy and diseased robusta coffee leaves. Expert plant pathologists analyzed the photographs after collecting them from various farms in Vietnam. Images of four distinct diseases, as well as healthy leaves and leaves with various kinds of damage, are all included in the dataset. Convolutional neural networks (CNNs) and other machine learning models' performance on the dataset was compared. With an accuracy of over 95% on the test set, they discovered that CNNs performed the best. The study makes a significant addition to the field of plant disease detection and emphasizes the need of creating extensive datasets for assessing machine learning-based techniques.

Ahmed et al [25] explain the four primary processes performed to get the dataset ready for mango leaf disease detection. Although many of these diseases emerge in the tree's leaves, learning about the frequent diseases that plague mango trees are the first step. The data is collected from four different mango orchards in Bangladesh and consists of healthy leaves along with seven diseases. The last phase involves manually classifying the photos in the dataset by humans with expertise, scaling the images to a standard form, and eliminating background noise from the images. It discusses the significance of data in machine learning and how best practices for dataset preparation must be adhered to in order to get the most out of machine learning models. In order to prepare the dataset, researchers must do a variety of processes, including representative samples, cleaning and enhancing and labelling the data.

A novel method for spotting diseases in tomato leaves is presented by Karthik, R., et al. [26] by combining attention mechanisms and residual convolutional neural networks (CNNs). to improve disease detection accuracy, an Attention Embedded Residual CNN (AER-CNN) framework incorporates attention mechanisms into the residual CNN architecture. The model achieved a high level of accuracy in identifying numerous diseases in tomato leaves, such as bacterial spot and early blight, after being trained on a dataset of tomato leaf images. The AER-CNN model outperformed conventional CNN models and other cutting-edge techniques.

Florian et al. [27] compare the categorization and detection of Esca disease in Bordeaux vines using two distinct feature extraction methods: Deep learning features as well as Scale-Invariant Feature Transform (SIFT) encoded features. The dataset comprises leaf pictures of both healthy and diseased vines for use in training and testing the algorithms. Deep learning features surpassed SIFT-encoded features and conventional machine learning formulas like Random Forest, SVM, and KNN in terms of accuracy. It suggests that the categorization and detection of Esca disease in vineyards may benefit from the application of deep learning characteristics.

The mechanism presented by López et al. [28] enables the assessment of uncertainty in a model's predictions. A Bayesian convolutional neural network (CNN) was employed by the scientists to identify diseases in images of several plant species' leaves. The model can provide a probability of each class and was trained using a dataset of labeled images, allowing for the quantification of uncertainty. When the Bayesian CNN model was compared to regular CNN models, it performed better and produced predictions that were more insightful. It implies that Bayesian deep learning, which offers both the prediction and the accompanying uncertainty, can be an effective way of detecting plant diseases.

Cristin [29] describes a method for detecting plant diseases that combines deep neural networks (DNNs) and the Rider-Cuckoo Search meta-heuristic optimization technique (RCS). The authors used RCS to optimize the hyperparameters of a DNN model trained on a dataset of images of various plant species' leaves. The images were then categorized as either healthy or unhealthy using the model. The findings demonstrated that the RCS-DNN model was highly accurate in identifying various plant diseases. The RCS-DNN model outperformed conventional DNN models and other cutting-edge techniques, based on the authors' findings. It implies that a very accurate and efficient method for detecting plant diseases can be achieved by combining RCS with DNN.

Oishi et al. [30] provide an automated method for spotting abnormal potato plants that makes use of portable video cameras and deep learning models. A portable video camera is used to take pictures of the potato plants, a deep learning model analyses the images, and a display system presents the results. In order to identify abnormal plants in the photos, a portable video-based device equipped with a convolutional neural network (CNN) model was deployed. It suggests that the proposed approach might work well in detecting abnormal potato plants automatically, with high accuracy, and in real time.

Ruchi et al. in [31] develops a CNN model to distinguish between healthy and pathological leaves on a dataset of leaf images from various plant species. The CNN model was then deployed to an embedded device like the Raspberry Pi, and it was successful in accurately detecting and identifying a number of plant leaf diseases in real-time. The device can be used as a reliable and reasonably priced way to find and recognize plant leaf diseases in the wild instantly.

Prasad et al [32] present a system that includes a mobile camera to captures images of leaf samples at different resolutions. After applying deep learning techniques to extract information from the images, the photos are classed as healthy or diseased using a combination of machine learning algorithms. The results showed that the system was quite good at identifying and categorizing different plant leaf diseases. When using multiple image resolutions, the approach performed best. Plant leaf diseases may be detected using the multi-resolution mobile vision system, which provides accurate results while being inexpensive and portable.

## 3. CASHEW DATASET DESCRIPTION

To maintain healthy cultivation, plant disease detection is essential to agricultural production since it continually monitors on and examines the health conditions of the plant during each stage of growth. There are currently many datasets that may be used right now that include images of different plants, including rice, paddy, wheat, tomatoes, etc. In contrast to short crops, the cashew tree's deep structure makes it challenging to identify the disease with ease. Hence, this dataset is created with the motive to provide a solution to the existing issue in cashew crops due to the lack of public datasets. There are two types of images in this dataset: "healthy" images and "anthracnose"-infected images. 1200 pictures taken using a mobile camera with a few from Google images with the natural background of the fields comprise the first dataset (DATASET 1). A part of the data is gathered from both online [33] and physical sources by photographing diseased and healthy cashew leaves on cashew farms. The images are 1080 x 1080 pixels in size. Recognizing the potential of CNN systems and the enormous number of parameters necessary to train them, it is discovered that the dataset's restricted size prevents it from being used to efficiently train a deep learning model, as it can impact its learning and predicting capabilities [34]. It can lead to overfitting or underfitting, lower model performance, and bias towards restricted data, resulting in incorrect predictions of new data and failure to represent the problem's complexity [35]. As a result, substantial volumes of diverse and representative data are required to adequately train deep learning models. To address the constraints of limited data approaches such as data augmentation and GAN are applied. The goal of this dataset's creation is to gather a range of images in various settings, including illumination, clarity, and quality, so that it may be used for more than just simple classification. Hence, DATASET 1 is expanded utilizing RGB color transformations, as well as geometric and photometric augmentations including rotation, zoom, shear change, etc., to produce DATASET 2, which consists of 5000 images. This second dataset is then used to train a GAN model. There might not be a wide range of various data on cashew leaves because these images only generate variations of the previously existing images. A third and final dataset (DATASET 3) is developed using RLSGAN to generate realistic synthetic images that have never existed before to overcome these limitations. The final dataset consists of images of original and augmented images of size 1080 x 1080 and GAN-produced images of size 64 x 64. The different types of datasets and their specifications are given below.

- DATASET 1: The first dataset contained 1200 images of plain cashew leaves and fruits, both healthy and diseased.
- DATASET 2: The second dataset included 5000 images of healthy and diseased cashew leaves and fruits that had been augmented geometrically and photometrically.
- DATASET 3: The third dataset comprised 20,127 images of healthy and diseased cashew leaves and fruits that were generated using GAN along with the original and augmented images

3.1 SPECIFICATION TABLE

Subject: Agriculture Engineering, and image processing, Data Science, Artificial Intelligence

| Specific subject area | Cashew leaves data sets for advanced anthracnose disease identification |
| --- | --- |

| Type of data | RGB Cashew leaves Image datasets |
|---|---|
| How the data were acquired | Hardware: Realme 3 Pro mobile phone with tripod stand<br><br>Software: Dedicated Python program<br><br>Each dataset is captured by following the standard protocols. |
| Data formats | .jpeg, .png |
| Description of data collection | The Cashew Leaves dataset is created using a Cashew farm, under natural environmental conditions to cover variations in illuminance, camera position, and occlusions, using an image-capturing device. The dataset comprises high-quality images of Cashew leaves captured in their natural habitat, providing a diverse range of images that can be used for various research and development purposes. This dataset can be particularly useful for tasks such as leaf segmentation, classification, and disease detection. |
| Data source location | Venue: District- Cuddalore, State-Tamil Nadu<br><br>Latitude and longitude of location: 11.610153579274524, 79.57183843400362<br><br>Temp: 20–30C°, Humidity: 60.0%, Annual Rain:1000-2000mm, Soil type: pure sandy soil |
| Data accessibility | Repository name: cashew leaves- healthy and anthracnose infected<br><br>Direct URL to data:<br><br>https://www.kaggle.com/datasets/varshajayaprakash/cashew-leaves-healthy-and-anthracnose-infected |
|  |  |

*3.2 VALUE OF DATASET*

- Currently, there is no publicly available dataset specifically designed for cashew leaves and focuses on anthracnose disease. This lack of a specialized dataset poses a significant challenge for researchers and developers working on tasks related to cashew leaf recognition, as it can limit the effectiveness and generalizability of their algorithms. To address this gap, we are creating a new dataset of cashew leaves captured in natural environmental conditions. The dataset will contain high-quality images of cashew leaves. With this new dataset, researchers and developers can train and evaluate their algorithms more effectively, leading to improved accuracy and robustness in cashew leaf recognition applications.
- The proposed dataset consists 3 datasets with 2 subcategories with leaves affected with anthracnose disease and healthy leaves and is generated using three methods 1 plain phone camera with 1200 images, 2 geometric and photometric augmentation with 5500 images 3 Generative adversarial network (GAN) with 20,127 images
- The quality of the images generated in the proposed dataset is optimized for use in real-time cashew leaf classification applications. Even low-quality drone cameras can be used to capture these images with sufficient clarity and resolution for accurate leaf classification. This makes the dataset especially useful for applications requiring low-cost, readily deployable imaging solutions, such as in agricultural settings or resource-constrained environments. The dataset is intended to facilitate the creation and testing of algorithms capable of performing real-time leaf classification using a range of imaging technologies and computational approaches.
- The proposed cashew leaf dataset can be immensely beneficial to researchers who are interested in agriculture and technology-based research, such as cashew leaf segmentation, classification, and disease detection using AI systems. With prior knowledge of deep learning and the Python computer language, it is simple for beginners in the field to start working from scratch using this dataset because it was produced in a structured pattern. It has a wide variety of high-quality images that can be used for various research and development tasks, giving researchers and developers plenty of chances to investigate various cashew leaf recognition apps. The dataset was created to make it easier to build and test algorithms for various tasks, including disease detection, segmentation, and classification of cashew leaves. It includes high-quality pictures of cashew leaves taken in their natural habitat. The dataset also only includes one variety of cashew leaves, but it can be expanded to include other types as well, allowing researchers to develop reliable recognition algorithms that work with various cashew varieties.

# 4. DATASET GENERATION METHODOLOGY

The methodology used to create the final dataset is a three-step process. It is more suitable for applications with limited availability of data and other resources. The method consists of several augmentation techniques for positional manipulations in data, a generative adversarial network (GAN) to expand the dataset by generating new images and to increase the authenticity and diversity of the data. Figure 1 shows the block diagram of the following procedure.

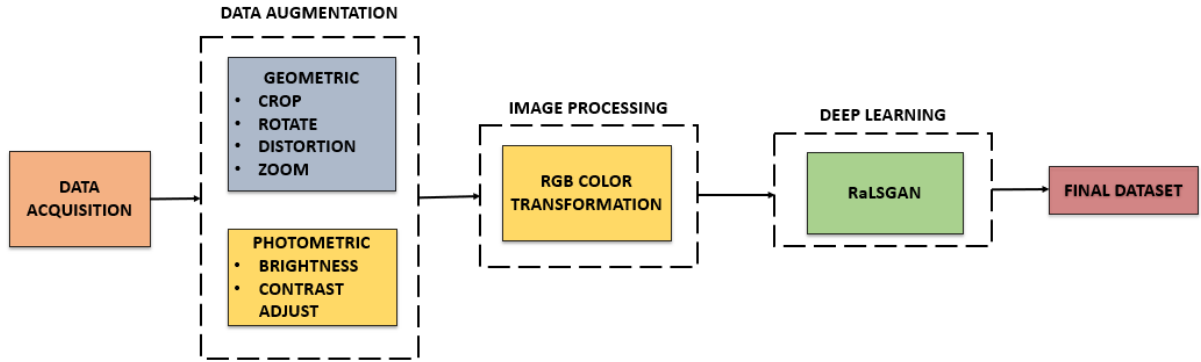

Figure 1. Cashew dataset generation methodology

*4.1 DATA AUGMENTATION*

The process of creating synthetic data by modifying specific properties of the original data is known as data augmentation. By artificially producing additional training samples from existing ones, data augmentation can assist eliminate bias in the data set and extend the diversity of images that the model can recognize. Augmentation is commonly used in computer vision and has been proven to improve model performance. [36]. Traditional augmentation methods are classified as geometric or photometric [37]. These strategies are applied at random to each picture in the training set, to create multiple copies of the same image to diversify the training data and reduce overfitting. In the first stage of our system, we apply simple geometric and photometric augmentation techniques like rotation, horizontal flip, zoom, random distortion, skew tilt, and color jitter. A separate pipeline is set up to generate around 5000 pictures and send them to the network's next level. This ensures that only the vector and pixel characteristics of the image are changed before the model is delivered to the GAN network. To further enhance the trainable parameters and maintain the uniformity of the images, a comparable but less intensive augmentation such as shearing, cropping, and scaling is carried out throughout the training stages of the classification model.

*4.2 IMAGE PROCESSING*

The image provided to the network to learn and train comprises a variety of information, including leaves, the diseased region, and some undesired background noise. Due to the absence of clear instructions, the model may be forced to train on incorrect data. Image processing is essential for computer vision because it prepares the raw images for analysis and interpretation by computer vision algorithms. Image processing techniques are used to improve picture quality, make features more visible, and extract relevant information from images [38]. RGB color transformation is used for this aim to emphasize the infected sections of the plant by transforming the color scheme of the images. The additive RGB color model combines different intensities of red, green, and blue light to create colors. Each pixel's color is a combination of these three basic colors when a picture is loaded in RGB format. The red, green, and blue channels [39] values can be modified to alter the color of each pixel. This enables many color adjustments and conversions, including modifying the image's brightness, contrast, saturation, and hue.

*4.3 RaLSGAN*

"GAN" stands for Generative Adversarial Network, a deep learning system built for generative tasks such as pictures, music, or voice generation [40]. A GAN works by training two neural networks, one termed the generator and the other the discriminator. The generator seeks to create artificial data that mimics real data, but the discriminator seeks to distinguish between the two. This procedure improves the discriminator's ability to identify fake data, while the generator increases its capacity to generate realistic data. GANs have been used in numerous areas, including computer vision, natural language processing, and audio processing. [41]. There are many GANs, each with a distinct set of abilities. Most GANs have the problem of not being easily capable of creating pictures since they demand very high computations and lack qualities relating to divergence reduction, prior knowledge, and gradient [42]. To overcome these limitations and produce high-quality images RaLSGAN is utilized. Figure 2 describes the working principle of GAN to generate images.

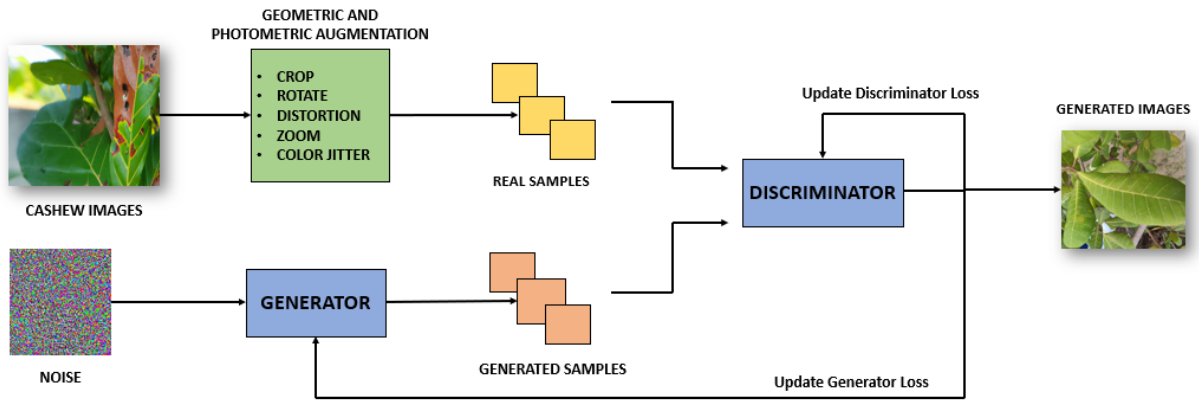

Figure 2: Working principle of GAN to generate images of cashew

RaLSGAN stands for Relativistic Average Least Squares Generative Adversarial Network, a version of the conventional GAN method. The loss function used to train the discriminator network is the primary distinction between a RaLSGAN and a standard GAN. The loss function in a classical GAN is commonly binary cross-entropy, which evaluates the discrepancy between the discriminator's forecasts and the actual data labels. The loss function in a RaLSGAN is a least squares loss that compares the discriminator's predictions to a relativistic average of real and fake data predictions and estimates the chance that real data is more realistic than either randomly chosen false data or fake data on average [43]. The primary reason for utilizing a RaLSGAN rather than a standard GAN is to stabilize the training process, increase the quality of produced samples, and provide high resolutions [44]. RaLSGANs have been demonstrated to be useful for several image production applications such as image super-resolution, style transfer, and class-conditional picture synthesis. The loss function of RaLSGAN is mentioned in equations (1) and (2).

$$disc\_loss = (C(x\_r) - AVG(C(x\_f)) - 1)^2 + (C(x\_f) - AVG(C(x\_r)) + 1)^2 \text{---------------(1)}$$

$$gen\_loss = (C(x\_r) - AVG(C(x\_f)) + 1)^2 + (C(x\_f) - AVG(C(x\_r)) - 1)^2 \text{---------------(2)}$$

The loss function for a generative adversarial network (GAN) is defined in terms of real and fake samples (x r and x f), their respective discriminator outputs (C (x r) and C(x f), and the average of those outputs for a set of samples (Avg(C(x r)). For a given set of samples, the discriminator's output and the average output are measured by two squared components that make up the loss function. The goal of adding or removing 1 is to achieve similar results for both true and false samples. With their roles reversed, the loss of the generator and discriminator are very similar. The generator trains one noise vector. Tanh activation is used in the final layer of several transposed convolutional layers with batch normalization and ReLU activation. The discriminator is similar in architecture with no final activation layer. A sigmoid function is used to predict the final output of the discriminator as either real or fake. The hyperparameters include the number of channels, kernel size =4, stride=2, and batch normalization settings for each layer, which may be modified depending on the application. With an Adam optimization strategy, the learning rate used is 0.001. During training, the loss of both the generator and the discriminator is tracked and stored. Images are generated continuously to show the model's progress.

## 4.4 DATASET CHARACTERISTICS

The cashew dataset has 20,127 pictures and is around 20 GB in size. There are two classes of images, with 81% of them being "healthy" and the remaining 19% being "anthracnose." 1200 images make up the first batch of data, which was gathered from google images and cashew farms. The second dataset has a total of 6700 images of size 1080 x 1080 which includes the original as well as augmented ones utilizing brightness, color jitter, zoom, rotate, and shear changes. The final dataset includes of photos that range in size from 1080 x 180 to 64 x 64, and in quality from low-quality GAN images to high-quality DSLR photographs. Also, varied heights, distances, and lighting situations, including both day and night, were used to collect the images for the dataset.

## 4.4 DATASET SPLITTING

To evaluate the classification performance of each dataset, testing and training sets are created for each dataset. An 80:20 split of the entire data set is used to create training and testing sets. The training folder contains a total of 16,101 images, of which 3,220 are anthracnose-infected and 12,880 are healthy. 4026 images, 3220 of which are in good health and 805 of which are anthracnose-infected, make up the remaining 20% of the testing set.

# 5. EVALUATION USING TRANSFER LEARNING

Since the model is to be implemented in real time in drones, it is necessary that the model be lightweight and take less time to train and make predictions. For this purpose, transfer learning is incorporated rather than training the model from scratch. Transfer learning is frequently used in computer vision to train deep neural networks for tasks like segmentation, object identification, and picture categorization. Several pre-trained models are utilized for this, including Inception and MobileNet. Large datasets like the ImageNet dataset, which has a diverse collection of annotated pictures, are used to train these algorithms. The pre-trained layers' weights are frozen to prevent them from being updated during training. To adapt the model to the new task, new layers are added. These layers are randomly initialized and trained to fit the new dataset [45]. The entire model is trained using the cashew images, including the pre-trained layers and the new layers. The models are fine-tuned by unfreezing certain layers to improve their performance. By reusing the pre-trained weights, the models can be quickly adapted to new tasks with smaller datasets. The last 30 layers of the base network are set to be trainable. The architecture of the model consists of the pre-trained feature extractor, followed by a flatten layer, a fully connected layer with 512 units and ReLU activation, a dropout layer with a rate of 0.2, and a final fully connected layer with softmax activation that produces the predicted class probabilities. A dense layer with a softmax activation function and a kernel regularizer with L2 regularization and a categorical cross-entropy loss function are also present in the output layer. The model is built with TensorFlow and Keras. The method employed 64 batches, 25 iterations, a learning rate of 0.01, a dropout rate of 0.2, and early stopping, which halts training if the validation loss does not decrease for three iterations in a row.

## 5.1 MOBILENET_V2

Embedded and mobile applications demand lightweight deep learning architectures that can be quickly deployed in resource constrained conditions without affecting computation effectiveness. Mobilenet V2 comprises a total of 53 layers, 32 of which are fully convolutional and the remaining 19 are residual bottleneck layers. By using a technique called depth wise separable convulsions to learn the relevant features from the images, the parameters and processing costs are reduced. [46]. The key difference between this model and a regular CNN is that it uses two types of filters, pointwise and depth wise. The depth wise component conducts a basic filtering operation on each channel of the input data, while the pointwise part uses a 1x1 convolution to combine the depth wise filters' output. As a result of this split, there are fewer parameters and faster processing. The residual block gets an input with extended channels and is eventually reduced back to its original size after several levels of computing [47]. This allows for the construction of larger networks with lower computational costs. After processing the input image and passing it through the depth wise separable convolutions and inverted residual blocks, the output is sent into the network's final layer, which performs the final classification.

## 5.2 INCEPTION_V3

Inception-v3 is a deep convolutional neural network with 48 layers. It was created to address the shortcomings of traditional convolutional neural networks (CNNs) by enabling more flexible and efficient filtering. In a CNN, an image model block known as an Inception Module attempts to approximate an ideal local sparse structure. Rather than being confined to utilizing only one type of filter size in a single image block, it allows us to use a choice of filter sizes, which we can then concatenate and pass onto the next layer.[48] A convolution seeks to decrease the dimensionality of the incoming data by using channel-wise pooling. The network's depth can thus be increased without risk of overfitting. The convolution in the channel dimension between each pixel of the picture and the filter is computed by a 1 X 1 convolution layer, also referred to as the bottleneck layer. The output will have a different number of output channels but the same height and breadth as the input. Convolutional filters of various sizes are used to aid the network in learning spatial patterns at various scales, just as the human eye does. The 3 X 3 convolution, as opposed to the 5 X 5 convolution, learns properties on a smaller scale.[49] Every inception architecture consists of the max-pooling layer seen in every neural network and a concatenation layer that joins the features acquired by the inception blocks. V3 is far more effective than earlier variants. To improve the model, the larger Convolutions were factored into smaller Convolutions.

## 6. RESULTS AND DISCUSSION

The CNN model underwent a testing process in stages using a diverse range of data, including both augmented images produced through geometric and photometric techniques and images generated through the use of GAN. The dataset utilized in this research was initially limited to 1200 images, which is comparatively small in scale. To ensure efficient computation using CNN, all images were resized to 64 x 64. The original images in DATASET 1 and their classes are presented in Figure 3.

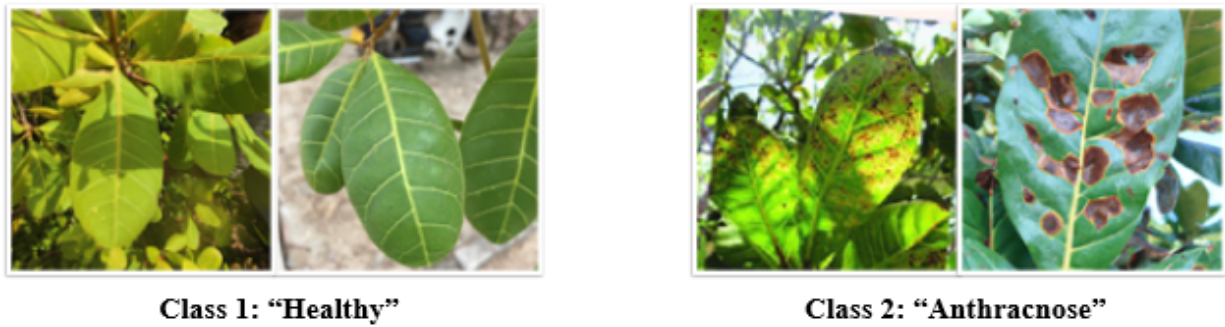

Figure 3: Images in DATASET 1 and their class names

Although it may suffice for certain use cases, DATASET 1 might not provide adequate data for deep learning models that necessitate a substantial amount of information to learn proficiently. To train the GAN effectively, a sufficiently large dataset is required to provide the generator network with enough variation and complexity to learn and generate realistic samples. If the dataset is too small, the generator network may not be able to learn the underlying distribution of the data effectively, resulting in poor-quality synthetic samples that do not accurately reflect the original dataset. In this study, to generate additional images for the GAN, augmentations such as rotation, horizontal flip, zoom, random distortion, and skew tilt are performed on the available dataset to generate more images of healthy and diseased cashew leaves and fruit. By doing so, it was possible to expand the dataset size (up to 5500 images ) beyond the original 1200 images and improve the quality of the dataset, thus providing the GAN with more data to learn from and improving its ability to generate realistic synthetic samples. Figure 4 depicts a sample of the augmented dataset (DATASET 2).

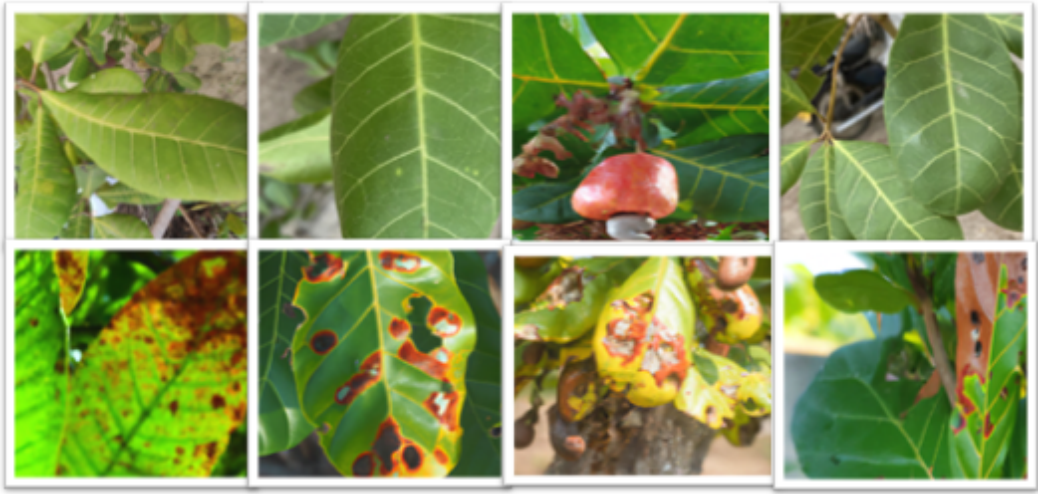

Figure 4: Geometrically augmentation images of cashew leaves and fruits in DATASET 2

The main goal of GAN is to increase the quantity and diversity of the training dataset in order to enhance the performance of the deep learning model. 5000 images of both healthy and infected leaves make up the training dataset, which has been enhanced to improve the dataset's variety. To evaluate the classification performance of each dataset, testing, and training sets are created for each dataset. Training and testing sets are split up in an 80:20 ratio. The images are then resized, normalized to a range of 0-1, and converted to tensor format to prepare them for training the GAN. The GAN is then trained for 300 epochs to generate around 14000 new, realistic images of both healthy and infected leaves to create DATASET 3. The produced images are of size 64 x 64 pixels and are depicted in Figure 5.

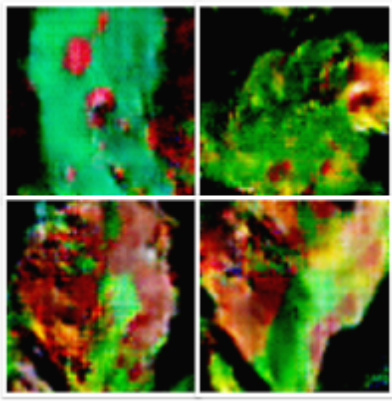

Figure 5: GAN-generated images

The performance of the deep learning model in real-time can be improved even though the quality of these produced images may not be as good as genuine photos. The main purpose of generating these images is to add more variety to the training dataset, which can help the model perform better in real-world scenarios and under more robust conditions Figure 6 shows the changes in the generator and discriminator loss over time since the model has been trained for a set number of epochs.

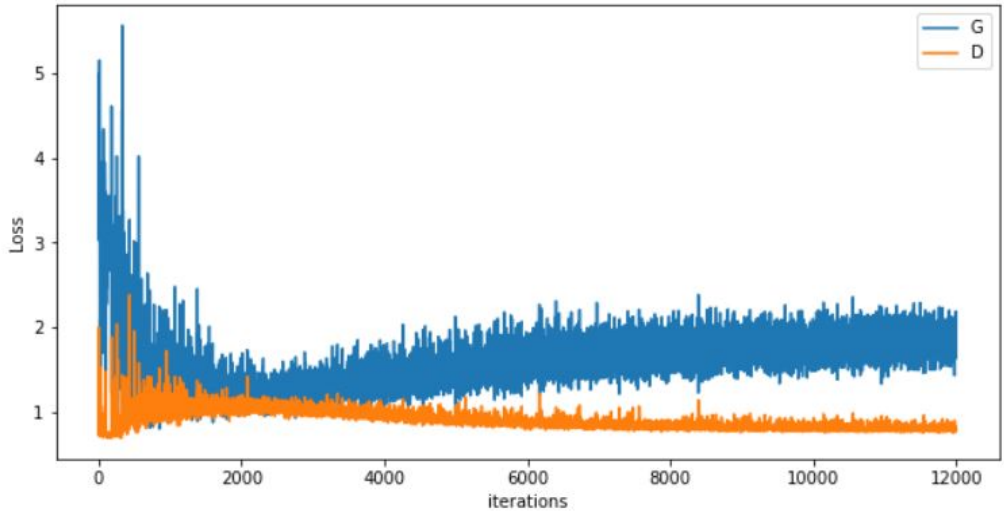

Figure 6: Loss curves of RaLSGAN model

Figure 6 shows that the GAN model achieves stability after 200 epochs and that the generator and discriminator loss do not change considerably. After reaching stability, the discriminator loss is approximately 0.7 and the generator loss is nearly 1.5. This shows that the GAN model is operating effectively, and that additional training after the stability phase may not result in significant improvements. This is due to the fact that the generator model is optimized to generate synthetic samples that are similar to real samples, whereas the discriminator model is optimized to differentiate between real and false samples. The models learn from each other as they are trained together, and the generator improves at producing realistic samples while the discriminator improves at recognizing fake samples. If the training is extended past the stability level, the models may begin to overfit to the training data, resulting in poor generalization performance. As a result, the generator may start producing unrealistic or low-quality samples, and the discriminator may become overly specialized to the training data, making it less effective at detecting fake samples. This ensures that the GAN model can generate high-quality synthetic samples that are comparable to real-world samples while remaining generalizable to new data.

As previously stated, the images are subjected to image processing such as brightness and contrast correction and RGB color transformation to highlight aspects such as diseased areas in the images. These image processing methods generally aim to increase the visibility and analyzability of the image's key components. The processed images are trained utilizing the trainable layers of the transfer learning network by including early stopping to track the validation loss, and the results obtained by MobileNet and Inception are compared to different types of datasets. Tables 1 and 2 provide a comparison of these outcomes.

Table 1: Comparison of results obtained using MobileNet_V2

| METHODOLOGY | ACCURACY % | | | NO. OF EPOCHS |
|---|---|---|---|---|
| | TRAINING | VALIDATION | TESTING | |
| DATASET 1 | 99.70 | 83.59 | 96 | 4 |
| DATASET 2 | 99.69 | 99.61 | 98 | 7 |
| DATASET 3 | 99.85 | 99.30 | 97.7 | 6 |

Table 2: Comparison of results obtained using Inception_V3

| METHODOLOGY | ACCURACY % | | | NO. OF EPOCHS |
|---|---|---|---|---|
| | TRAINING | VALIDATION | TESTING | |
| DATASET 1 | 98.63 | 84.38 | 83.2 | 7 |
| DATASET 2 | 99.19 | 98.96 | 98.6 | 12 |
| DATASET 3 | 99.30 | 99.68 | 95.3 | 8 |

For successful training, the diseased and healthy images are divided in a 20:80 ratio. Since the initial collection of images contains just 1200 images, obtaining accurate results is challenging. As a result, the typical validation and testing accuracies for Mobilenet and Inception are fairly low, averaging around

84%. Transfer learning has helped to reduce training time and the number of epochs. In this application it is observed that MobileNet has a lower training time and slightly higher accuracy than Inception due to the presence of depth wise separable convolutions in the former that makes it light and less computationally intensive. After implementing GAN and evaluating the model with those results using 20,127 images, it appears that the network learns successfully from a range of inputs while producing high accuracy results. MobilNet once again outperforms Inception in terms of testing accuracy and training time, achieving 97.7% accuracy in just 6 epochs as opposed to Inception's 95.3% accuracy in 8 epochs.

Since the network structure is now defined, the naïve models are changed to TensorFlow Lite in order to function effectively at the hardware level. To improve the model and decrease computing constraints during inference time, the original network is fused. It is done to minimize latency, memory, and inference time with minimum changes in classification accuracy. The initial model is first converted to float32 format and later converted to other types. Nevertheless, post training quantization is utilized to further reduce memory and give the model the ability to work in many contexts. We employ float 16 quantization, complete integer quantization, and dynamic range quantization. In CPU-based systems, dynamic range quantization enhances model performance by reducing memory by four times and boosting speed by two to three times. Full integer quantization provides the same benefits but also makes the model compatible with GPU, Edge TPU, and Microcontrollers in addition to CPU. The results obtained post quantization is given in Figure 7.

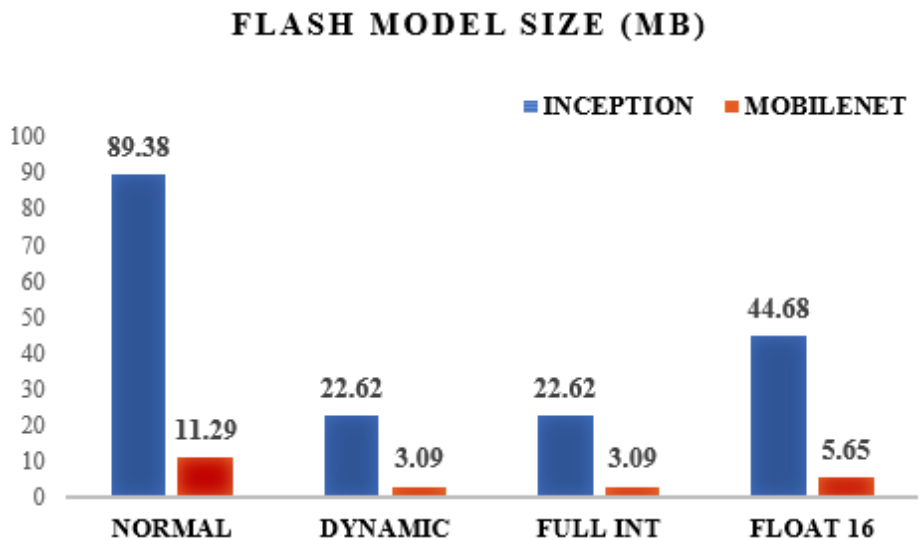

Figure 7: Comparison of model size

It can be inferred that standard conversion to Tflite takes up the most memory in both architectures. As previously indicated, the findings show that dynamic and full integer optimization reduces memory by four times and float 16 by two times. Mobilenet once again outperforms Inception in terms of flash model size, with 87% less RAM consumption. Because of memory and other resource limits, the larger file size in Inception after conversion and quantization may not be effective at the hardware level. Mobilenet, on the other hand, took up just 3.09 MB, making it suitable for use in edge devices like microcontrollers. The improved MobileNet models were analyzed on a simulator using the MLTK profiler [50]. The calculations are based on the ARM Cortex-M33 simulation environment with an MVP accelerator running at 78 MHz On assessment, it is discovered that MobileNet architecture consumes around 6 MB of run time memory, has a latency of 10.7 s, and an average inference time of 93.8 ms.

## 7. FUTURE SCOPE

The integration of image processing and classification algorithms into the drone's software is an important next step in implementing the suggested model on a drone. The drone should be equipped with a high-definition camera capable of capturing the cashew harvest from an appropriate height and angle. Furthermore, the drone should be equipped with a precision spray mechanism to guarantee that the pesticide spray is only directed towards infected regions. To assess the model's real-time accuracy in the cashew field, the system should be tested in a variety of weather and light conditions. Field experiments should mimic real-world settings and involve a variety of situations, such as cashew tree size and form changes. This assists in identifying potential system limits or areas for improvement in system accuracy.

Another area for development is to make the datasets for training the deep learning model more robust. The current dataset contains images collected from a single field under various weather and lighting conditions, but it may be useful to expand the dataset to include images from different regions, fields, and seasons. This will improve the model's generalization for a new environment and help to increase the accuracy in practical applications. Furthermore, the suggested approach may be used to various crops and diseases in order to boost agricultural productivity while reducing chemical use. The approach, for example, may be extended to identify diseases in other tree crops like mango and avocado. This will aid in addressing broader agricultural challenges and contributing to long-term food production. Overall, the proposed model and drone-based system represent a promising method for improving disease detection and treatment in cashew plants. Further work may be done to increase the system's efficiency, usefulness, and sustainability, as well as to realize its potential uses in diverse agricultural situations.

## 8. CONCLUSION:

This paper presents a dataset that contains 20,127 images of healthy and anthracnose-infected cashew leaves and describes the use of TinyML to develop an intelligent cashew disease detection system. Due in part to a shortage of diverse data, multiple procedures such as geometric and photometric augmentation are performed, followed by RaLSGAN to produce diverse realistic data to broaden the dataset's diversity. Because the model is to be implemented in real time in edge devices, a transfer learning technique is used to build a binary classifier to evaluate the efficiency of the dataset using the MobileNet and Inception architectures. To minimize memory needs, trained models are transformed to tflite files and subjected to post-training optimization. The models perform well, with an accuracy of 97.7% when using MobileNet and 95% when using Inception. Moreover, Mobilnet appears to take extremely little memory after quantization, making it an excellent alternative for deployment in edge devices. The model needs an average inference time of 93.8 ms and has a latency of 10.7s when tested on an ARM Cortex M33 at 78 MHz.